\documentclass[twocolumn,showpacs,preprintnumbers,amsmath,amssymb]{revtex4}
\usepackage{tabularx,graphicx}

\usepackage{color}
\usepackage{hyperref}
\hypersetup{
    colorlinks=true,
    linkcolor=blue,
    filecolor=blue,      
    urlcolor=blue,
}

\usepackage{color} 
\newcommand{\itt}{\it}

\usepackage{ulem}   

\begin{document}

\newcommand{\beq}{\begin{equation}}
\newcommand{\eeq}{\end{equation}}
\newcommand{\beqn}{\begin{eqnarray}}
\newcommand{\eeqn}{\end{eqnarray}}
\newcommand{\bmath}{\begin{subequations}}
\newcommand{\emath}{\end{subequations}}
\newcommand{\bra}[1]{\langle #1|}
\newcommand{\ket}[1]{|#1\rangle}

\title{Nonstandard  superconductivity or no superconductivity in hydrides under high pressure}

\author{J. E. Hirsch$^{a}$  and F. Marsiglio$^{b}$ }
\address{$^{a}$Department of Physics, University of California, San Diego,
La Jolla, CA 92093-0319\\
$^{b}$Department of Physics, University of Alberta, Edmonton,
Alberta, Canada T6G 2E1}

\begin{abstract} 
Over the past six years, superconductivity at high temperatures has been reported in a variety of hydrogen-rich compounds
under high pressure. That high temperature superconductivity should exist in these materials is expected according to the
conventional theory of superconductivity, as shown by detailed calculations. However here we argue that
 experimental observations rule out conventional superconductivity in these materials.
Our results indicate  that either these materials are unconventional superconductors of a novel kind, which we term
`nonstandard superconductors',
or alternatively that  they are not superconductors. If the former, we point out that the critical current in these materials should be
several orders of magnitude larger than in standard superconductors, potentially opening up the way to important technological applications. If the latter, 
which we believe is more likely, we suggest that the signals interpreted as superconductivity
are either experimental artifacts or they signal other interesting physics but not superconductivity.

\end{abstract}
\maketitle 

\section{introduction}
According to the conventional BCS-Eliashberg theory of superconductivity \cite{eliash}, superconductivity at high temperatures should 
exist in materials with light ions and
large electron-phonon coupling. Neil Ashcroft suggested in 1968 \cite{ashcroft} that such conditions would be met in hydrogen turned metallic at
high pressures, and in 2004 extended his prediction to include metal hydrides  with high hydrogen content \cite{ashcroft2}. Motivated by
these theoretical predictions, substantial experimental and theoretical efforts have been expended in recent years searching for high   temperature 
conventional superconductivity in metal hydrides under high pressure \cite{review1,review2,review3}. Beginning in 2015 with Eremets et al's discovery of
superconductivity in pressurized hydrogen sulphide  \cite{eremetssh}, several metal hydrides under high 
pressure have been found to exhibit high temperature superconductivity \cite{eremetssh,eremetsp,eremetslah,hemleylah,yttrium,yttrium2,thorium,roomt,layh10}, culminating in the recently reported discovery of room temperature
superconductivity in a compound with carbon, sulphur and hydrogen \cite{roomt} (hereafter called CSH).

Together with these experimental discoveries, sometimes before and sometimes shortly thereafter, a variety of detailed calculations have been
performed based on BCS-Eliashberg theory  that both predicted and apparently confirmed that these
materials are high temperature conventional superconductors \cite{th0,th1,th2,th3,th4,th5,th6,th7,th8,th9,th10}. 
Sometimes superconductivity was `confirmed' without the need of $any$  experimental evidence  \cite{th11,th12,ceh9,dias2}. 
It should be pointed out however that these calculations are applied in a range of electron-phonon coupling constant $\lambda\gtrsim 2$ where
numerical calculations \cite{lambda1} and theoretical considerations \cite{lambda2,lambda3,lambda4} suggest that Eliashberg theory is no longer applicable 
due to polaron formation. In addition all these calculations rely
on estimating the `Coulomb pseudopotential' $\mu^*$ that cannot be calculated independently, that can 
suppress superconductivity if it is sufficiently large \cite{eliash,webb}.
Furthermore, these calculations do not take into account the possibility of competing instabilities such 
as charge density wave \cite{cdw,marsiglio90} and magnetic phases \cite{magnetic,fm,mazov}.
In this paper we show   that in fact {\it these materials cannot be conventional
superconductors}, contrary to the experimental and theoretical evidence cited above.

Superconducting materials \cite{specialissue} are generally understood to be either `conventional' or `unconventional' \cite{unconv}.  
Unconventional superconductors are materials  that do not
become superconducting driven by the BCS-electron-phonon interaction but rather through some different physics, 
generally believed to be related to   strong electron-electron interactions, for example the high $T_c$ cuprates \cite{cuprates} and iron pnictides \cite{pnictides}. 
However even unconventional superconductors exhibit  standard  properties of superconductivity such as the Meissner effect,
 upper and lower critical fields, critical current, etc \cite{tinkham}.  
  In this paper we call such superconducting materials \cite{specialissue}
 `standard superconductors' and
 we introduce a new category of non-conventional superconductors, which we call {\it `nonstandard superconductors'}: materials that do not
 exhibit basic properties of superconductors that standard superconductors, both conventional and unconventional, share. 
 We will argue
 that hydrides under pressure, $if$ they are superconductors,  belong to this new category of superconductors. 
  
According to the established understanding of superconductivity \cite{tinkham}, there are three lengths that play a critical role in
 all standard superconductors, whether conventional or unconventional: (i) the London penetration depth $\lambda_L$, the Pippard coherence length $\xi$, and the
 electron mean free path $\ell$. Type I superconductors have $\lambda_L<\xi$ and are a limited set of materials, mostly
 elements in pure form, all believed to be conventional superconductors \cite{webb}. Most superconducting materials, both conventional and unconventional, are type II, with $\xi<\lambda_L$. 
 A strongly type II material has $\xi <<\lambda$. Disorder  (short $\ell$) 
 increases $\lambda_L$ and turns a type I material into type II. Most standard superconductors, and in particular the
unconventional cuprates and iron pnictides, are type II materials. The higher the $T_c$ the more strongly 
 type II the material tends to be. 
 
 In this paper we argue that hydride superconductors exhibit properties characteristic of {\it both} type I and
 type II superconductors, in particular strongly type II. Standard superconductors can be borderline type I/ type II 
 if $\xi \sim \lambda_L$ but 
 they cannot be type I and strongly type II simultaneously. Therefore, we will argue that  hydrides under pressure are
 `nonstandard superconductors' or, alternatively, that they are not superconductors.
 
 A preliminary account of this work can be found in Ref. \cite{hm}. 
 While the present  paper was in the final stages of completion, a paper by Dogan and Cohen appeared \cite{cohen} that  expanded on our
 arguments  \cite{hm} that the behavior of  CSH   \cite{roomt}
is incompatible with standard superconductivity and independently suggested the possibility of nonstandard superconductivity.
 
\section{type I vs type II superconductivity}

As summarized in Sect. II of \cite{webb}, the phenomenology of the response of standard superconductors to
magnetic fields is common to conventional and unconventional, i.e. standard, superconductors. 
A microscopic theory is not required to describe this response \cite{tinkham}.

        \begin{figure} []
 \resizebox{8.5cm}{!}{\includegraphics[width=6cm]{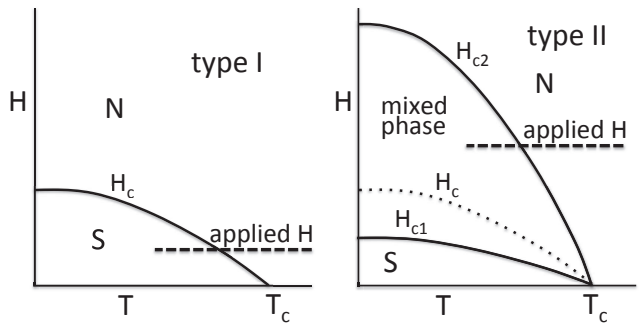}} 
 \caption {Schematic phase diagrams of type I and type II superconductors in a magnetic field.
 N=normal state, S=superconducting state.}
 \label{figure1}
 \end{figure} 

Fig. 1 shows schematically the phase diagrams of type I and type II superconductors
when a magnetic field H is applied.  In a type-I superconductor
the magnetic response is perfect diamagnetism: 
the magnetic field is completely expelled (except within a region of thickness $\lambda_L$
from the surface) provided the field strength
is less than a critical value $H_c(T)$, the thermodynamic critical field. Above this field value the material
is in the normal state. When the system is cooled in an applied field H,
it undergoes a reversible  first order phase transition from the normal to the superconducting state
at the temperature $T(H)$ where $H=H_c(T(H))$. 
The resistance of the material is finite for $T>T(H)$ and drops to zero 
discontinuously at $T=T(H)$ in an ideal situation. In the presence of imperfections in the
sample the transition will be slightly broadened both in the absence and in the presence
of applied magnetic field.

Instead, in a type-II superconductor the
material exhibits perfect diamagnetism only up to a critical field $H_{c1}$ smaller than $H_c$;
with increasing applied field, flux begins to penetrate the material
in the form of vortices. This continues to occur up to an upper
critical field, $H_{c2}$, above which the material  becomes normal.
When cooling a type II superconductor in an applied field $H$, the system will undergo
a second order phase transition at the temperature $T(H)$ where 
$H=H_{c2}(T(H))$ where  it enters the `mixed phase', with vortices in its interior,
their density depending on temperature and magnetic field.
In this mixed phase, whether or not the system has finite or zero resistance depends
on the behavior of the vortices, whether or not they are `pinned'. We will discuss this in a later section. 
Note from Fig. 1 that upon lowering the temperature for given applied $H$ the system
necessarily enters the mixed phase first. If the applied field is sufficiently
small so that $H<H_{c1}(T=0)$, at a lower temperature the system will enter the
perfectly conducting state with zero resistance. Instead, if
$H>H_{c1}(T=0)$ the system remains in the mixed phase down to zero temperature.

Whether a material is a type-I or a type-II superconductor
depends on the  Ginzburg-Landau parameter $\kappa=\lambda_L(T=0)/\xi(T=0)$. Since the penetration depth increases
as the mean free path decreases \cite{tinkham}
 a type-I superconductor
can be made into a type-II superconductor through disorder.
The thermodynamic critical field $H_c(T)$  is
determined by the difference in the free energy per unit volume between normal and
superconducting states as
\beq
f_n(T)-f_s(T)=\frac{H_c(T)^2}{8\pi}
\eeq
both for type I and type II superconductors,
and at zero temperature it is given within BCS theory by
\beq
\frac{H_c(0)^2}{8\pi}= \frac{1}{2} g(\epsilon _F) \Delta (0)^2
\eeq
where $\Delta$ is the energy gap and $g(\epsilon_F)$ the density of states at the Fermi energy.
Within Ginzburg-Landau (GL) theory the thermodynamic critical field is given by
\beq
H_c(T)=\frac{\phi_0}{2\sqrt{2}\pi \lambda_L (T)\xi(T)}
\eeq
where $\phi_0=hc/2e$ is the flux quantum.
The lower and upper
critical fields for type II superconductors are given within GL theory by
\bmath
\beq
H_{c1}(T)=\frac{\phi_0}{4\pi\lambda_L(T)^2} ln(\kappa(T))
\eeq
with $\kappa(T)=\lambda_L(T)/\xi(T)$,
\beq
H_{c2}(T)=\frac{\phi_0}{2\pi \xi (T)^2}
\eeq
\emath and are related to the thermodynamic critical field by
\beq
H_{c1}H_{c2}=H_c^2 ln(\kappa) .
\eeq
A strongly type II superconductor has $\kappa>>1$ and hence
$H_{c1}<<H_c<<H_{c2}$.

To get simple explicit expressions for the temperature dependence we can use the two-fluid model. The thermodynamic critical field is given by
 ($t \equiv T/T_c$)
\beq
H_c(T)=H_c(0)(1-t^2).
\label{hc_gl}
\eeq
and the London penetration depth by
\beq
\lambda^2_L(T)=\lambda^2_L(0)\frac{1}{1-t^4}.
\eeq
The upper critical field is then given by
\beq
H_{c2}(T)=H_{c2}(0)\frac{1-t^2}{1+t^2}
\eeq
and the coherence length by
\beq
\xi(T)=\xi(0)\sqrt{1+t^2 \over 1-t^2}. 
\label{xi_gl}
\eeq

\section{are hydride superconductors type I or type II?}
We argue that experimental evidence clearly shows that hydride superconductors,
particularly those with high $T_c$, are strongly type II. Ideally, one would want
to measure both $\xi$ and $\lambda_L$ to answer this question. However in none of
the experiments reported so far was  $\lambda_L$ inferred from experiment
(in \cite{roomt} an erroneous value for $\lambda_L$ not based on experimental
evidence was given \cite{hm,cohen}). Experimental values for the coherence length $\xi$ are inferred
from the lowering of the transition temperature under an applied magnetic field.
Here we argue that those reported results and theoretical considerations necessarily
imply that $\lambda_L>>\xi$, hence that the materials are strongly type II
(if they are superconductors). 

Consider the carbonaceous sulfur hydride (CSH) recently reported to be a room temperature
superconductor. Resistive transitions in applied magnetic field 
are shown in Fig.~2 of Ref.~\cite{roomt}, showing that the transition temperature drops from $287$ K 
for no applied magnetic field to increasingly lower values as the applied magnetic field increases. 
Let us consider the possibility that CSH might be a type I superconductor. 
The temperature dependence of the thermodynamic critical field 
is given within the two-fluid model by Eq.~(6).
In an applied field of $9$ T, the resistance drops almost discontinuously (in a range $\Delta T <1K)$
from a finite value to essentially zero \cite{roomt}. This then implies that
$H_c(265K)=9T$. 
With  $T_c=287K$ Eq. (6)  yields $H_c(T=0)=61T$. Within BCS theory
the energy gap at zero temperature and the  critical temperature are related by
\beq
\frac{2\Delta(0)}{k_B T_c}=3.53
\eeq
yielding $\Delta(0)=44 \ {\rm meV}$.
From Eq. (2) this yields for the density of states
at the Fermi energy
\beq
g(\epsilon_F)=\frac{9.7 states}{eV \AA^3} .
\eeq
Assuming the density of states is given by the free electron expression with an effective
mass $m^*$ yields
\beq
g(\epsilon_F)=(\frac{3}{\pi^4})^{1/3} n^{1/3} \frac{m^*}{\hbar^2}=\frac{0.0411}{eV\AA^2}n^{1/3}(\frac{m^*}{m_e})
\eeq
with $n$ the number density and $m_e$ the bare electron mass. The band effective mass should be close to the bare electron
mass, corrected by the mass enhancement resulting from the electron-phonon interaction.
Assuming an electron-phonon coupling constant $\lambda\sim 2$ resulting from 
Eliashberg calculations yields
\beq
m^*=m_e(1+\lambda)\sim 3m_e
\eeq
yielding
\beq
g(\epsilon_F)=\frac{0.123}{eV \AA^2}n^{1/3} .
\eeq
The Wigner-Seitz radius (radius of a sphere with volume equal to the volume per
conduction electron) is
\beq
r_s=(\frac{3}{4\pi n})^{1/3}
\eeq
so that in order for the density of states Eq. (14) to give the value
Eq. (11) requires
\beq
r_s=0.0079 \ \AA
\eeq
which is   much smaller than half of any interatomic distance in CSH or any other hydride under pressure \cite{th11}. 
This establishes that CSH cannot be a type I superconductor. The same applies to all
other reported superconducting hydrides, since invariably  applied magnetic fields of several Tesla give small shifts in the critical temperature.
We note that Talantsev and coworkers  have also 
concluded that sulphur hydride and other hydrides are strongly type II  \cite{tsal,tsal2}, and Dogan and Cohen concluded the same for CSH \cite{cohen}.

We conclude then that hydrides under pressure, if they are superconductors, are type II
superconductors. The reduction in $T_c$ with applied magnetic field then reflects 
the upper critical field $H_{c2}$ and not the thermodynamic critical field $H_c$.
Under this assumption, 
for the case of CSH, extrapolation to zero temperature 
using either GL theory or other theoretical approaches yields 
for the zero temperature upper critical field values
in the range $H_{c2}(0)=50T$ to $85T$ \cite{roomt}. Assuming the value obtained from GL theory
$H_{c2}(T=0)=62T$ yields from Eq. (4b) a zero temperature coherence length $\xi(T=0)=23\AA$.
Similar values for $\xi$ are obtained for the other superconducting hydrides.

For type I superconductors, coherence lengths are at minimum several hundred $\AA$.
Coherence lengths of order $20\AA$ indicate the materials are strongly type II.
For example, for the conventional superconductor 
$MgB_2$, $\xi\sim 50\AA$ and $\lambda_L\sim 1400 \AA$. For YBCO, $\xi \sim 30 \AA$ and $\lambda_L \sim 1500 \AA$.
For a rough estimate of $\lambda_L$ here, we use the BCS formula for the coherence length \cite{tinkham}
\beq
\xi=\frac{\hbar v_F}{\Delta}=\frac{\hbar k_F}{m^* \Delta}
\eeq
together with the relation
\beq
k_F=(3\pi^2 n_s)^{1/3}
\eeq
and the London penetration depth given by \cite{tinkham}
\beq
\lambda_L=\sqrt{m^*c^2/4\pi n_s e^2}
\eeq
to yield
\beqn
\lambda_L&=&\sqrt{\frac{3(\hbar c)^6}{4\pi^2 e^2 \Delta^3 (m_e c^2)^2 \xi_0^3}}(\frac{m_e}{m^*}) \nonumber \\
&=&1935(\frac{1}{\Delta(eV)\xi(\AA)})^3(\frac{m_e}{m^*}) .
\eeqn
In particular for CSH with $\xi=23\AA$ and $\Delta(0)=0.0437 eV$,
\beq
\lambda_L=1906 \AA (\frac{m_e}{m^*})
\eeq
so that $\lambda_L>>\xi$ for any reasonable value of $m_e/m^*$.

\section{ resistive broadening in standard conventional and unconventional 
superconductors}

In standard conventional and unconventional type II superconductors it is empirically
observed that the resistive transition is broadened in an applied magnetic field,
the more so the larger the field. It is also observed that the broadening is
larger for materials with higher $T_c$, which are also strongly type II ($\lambda_L>>\xi$).
Typical data for broadening of the resistive transition in standard superconductors are shown in Figs. 2-5 for  $Mg B_2$, 
$NbN$ and $Mo_xRe_{1-x}$ (conventional)
and $YBCO$ (unconventional) respectively.  

As discussed in Sect. II, when a type II superconductor is cooled in a magnetic field $H$, it first enters the mixed state, with vortices
in its interior. A vortex has a normal core of radius $\xi$ and carries magnetic flux $\phi_0$, the flux quantum.
Right at $T_c(H)$ (defined as $H=H_{c2}(T_c(H))$) the number of vortices per unit area is $1/(2\pi \xi^2)$, so that the
vortex cores of radius $\xi$, where the system is normal, almost overlap. As the temperature is further lowered, with constant applied $H$,
the density of vortices stays roughly constant but their core size shrinks, as more electrons condense into the superconducting state.

Figures 2 to 5 show that as the system enters this mixed state, the resistivity does not drop discontinuously to zero but rather gradually.
So there is dissipation as current flows through the mixed state. As the temperature is further lowered, the resistivity drops to zero
over a temperature range $\Delta T$ that is an increasing function of the applied magnetic field $H$. 
This is the expected behavior in all standard type II superconductors, which can be understood theoretically using concepts
developed more than 50 years ago, that apply to both conventional and unconventional superconductors \cite{tinkham}. We review this
theory in a following section.
        \begin{figure} []
 \resizebox{6.5cm}{!}{\includegraphics[width=6cm]{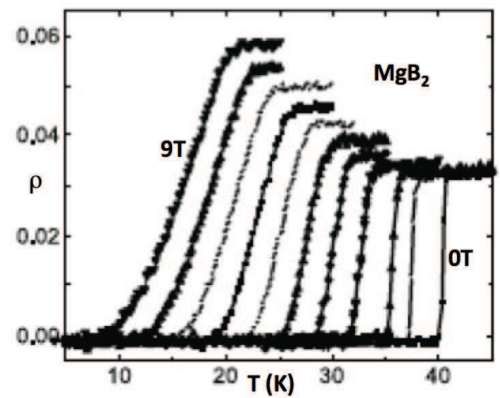}} 
 \caption {Resistive transition for $MgB_2$ in a magnetic field ranging from 0 to 9T, from
  P.C. Canfield, S.L. Bud'ko and D.K. Finnemore, \href{https://www.sciencedirect.com/science/article/pii/S0921453402023286}
     {Physica C: 385, 1 (2003)}, Ref. \cite{canfield}.}
 \label{figure2}
 \end{figure} 
 
         \begin{figure} []
 \resizebox{8.0cm}{!}{\includegraphics[width=6cm]{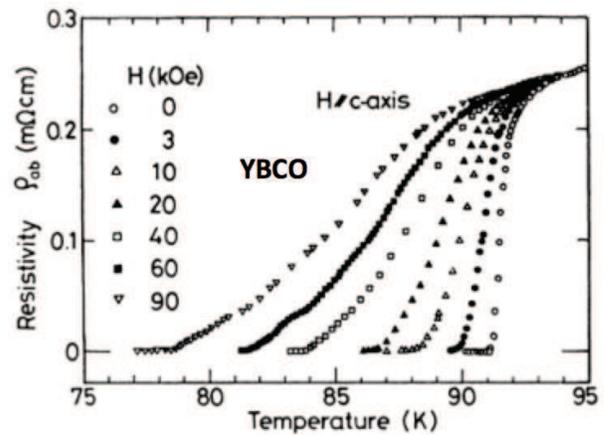}} 
 \caption {Resistive transition for $YBCO$ in a magnetic field ranging from 0 to 9T, from
 Y. Iye, T. Tamegai, H. Takeya, and H. Takei, in Superconducting
Materials, edited by S. Nakajima and H. Fukuyama,
Jpn. J. Appl. Phys. Series I (Publication Office, Japanese Journal
of Applied Physics, Tokyo, 1988), p. 46, Ref. \cite{iye}.}
 \label{figure3}
 \end{figure} 
 
         \begin{figure} []
 \resizebox{8.0cm}{!}{\includegraphics[width=6cm]{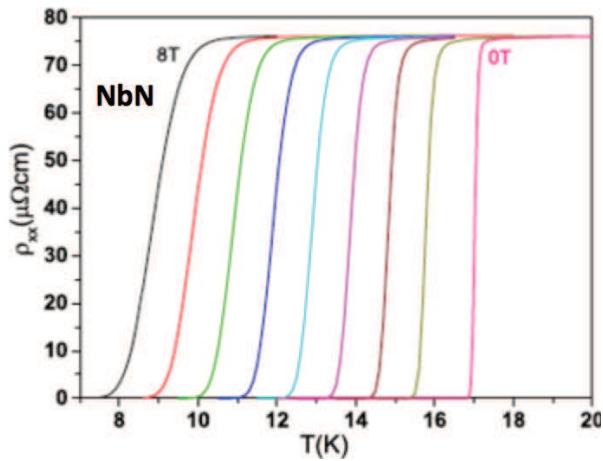}} 
 \caption {Resistive transition for $NbN$   in a magnetic field ranging from 0 to 8T, from
 D. Hazra, N. Tsavdaris, S. Jebari, A. Grimm, F. Blanchet, F. Mercier, E. Blanquet, C. Chapelier and M. Hofheinz,
 \href{https://iopscience.iop.org/article/10.1088/0953-2048/29/10/105011/meta}{ Sup. Sci. and Tech. 29,  10501 (2016)}, Ref. \cite{nbn}.}
 \label{figure4}
 \end{figure} 
 
       \begin{figure} []
 \resizebox{8.0cm}{!}{\includegraphics[width=6cm]{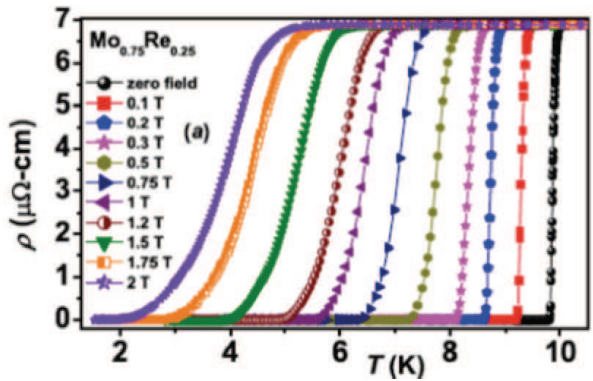}} 
 \caption {Resistive transition for $Mo-Re$ alloy in a magnetic field ranging from 0 to 2T, from
 S. Sundar, M. K. Chattopadhyay, L.~S. Sharath Chandra, R. Rawat and S. B. Roy,
  \href{https://iopscience.iop.org/article/10.1088/0953-2048/30/2/025003}
 { Sup. Sci. and Tech. 30,  025003 (2017)}, Ref. \cite{more}.}
 \label{figure5}
 \end{figure}

\section{expected versus observed broadening of the resistive transition in hydrides}

       \begin{figure} [t]
 \resizebox{6.5cm}{!}{\includegraphics[width=6cm]{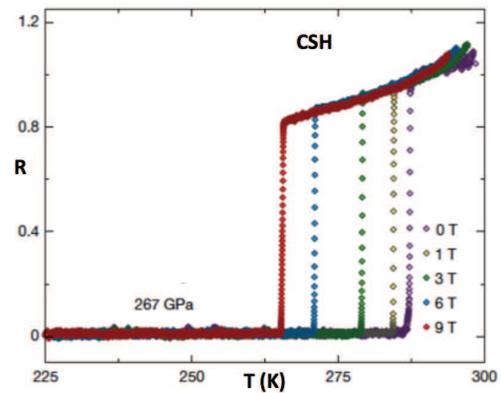}} 
 \caption {Resistive transition for $CSH$ in a magnetic field ranging from 0 to 9T, from,
Elliot Snider, Nathan Dasenbrock-Gammon, Raymond McBride, Mathew Debessai, Hiranya Vindana, 
     Kevin Vencatasamy, Keith V. Lawler, Ashkan Salamat and Ranga P. Dias,  
     \href{https://www.nature.com/articles/s41586-020-2801-z}{Nature 586, 373 (2020)}, Ref. \cite{roomt}.}
 \label{figure6}
 \end{figure} 
 
        \begin{figure} [t]
 \resizebox{6.5cm}{!}{\includegraphics[width=6cm]{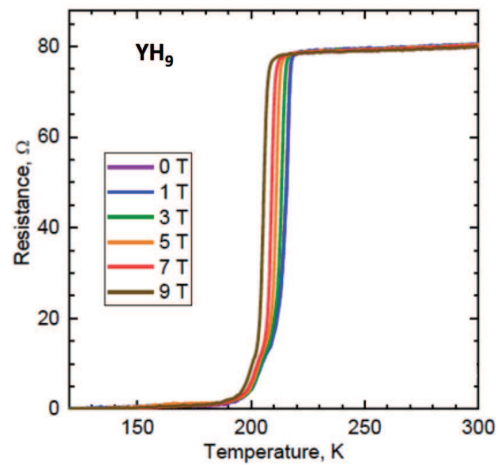}} 
 \caption {Resistive transition for $YH_9$ in a magnetic field ranging from 0 to 9T, from
P. P. Kong, V. S. Minkov, M. A. Kuzovnikov, S. P. Besedin, A. P. Drozdov, S. Mozaffari, L. Balicas, F.F. Balakirev, V. B. Prakapenka, E. Greenberg, D. A. Knyazev, M. I. Eremets,  \href{https://arxiv.org/abs/1909.10482}{arXiv:1909.10482 (2019)}, Ref. \cite{yttrium2}.}
 \label{figure7}
 \end{figure} 
 
        \begin{figure} [t]
 \resizebox{5.5cm}{!}{\includegraphics[width=6cm]{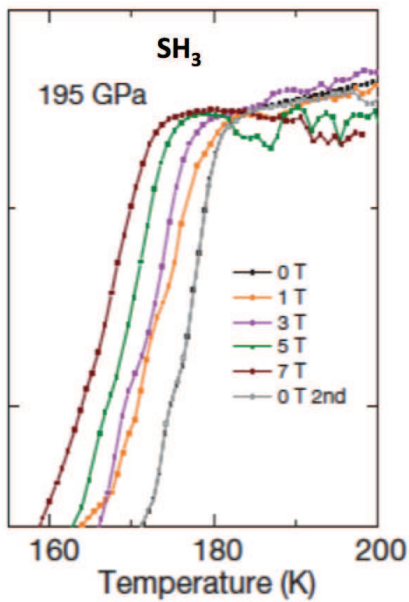}} 
 \caption {Resistive transition for $SH_3$ in a magnetic field ranging from 0 to 7T, from
 A.P. Drozdov, M.I. Eremets, I.A. Troyan, V. Ksenofontov and S.I. Shylin,
   \href{https://www.nature.com/articles/nature14964}{Nature 525, 73-76 (2015)}, Ref. \cite{eremetssh}.}
 \label{figure8}
 \end{figure} 
 
        \begin{figure} [t]
 \resizebox{6.5cm}{!}{\includegraphics[width=6cm]{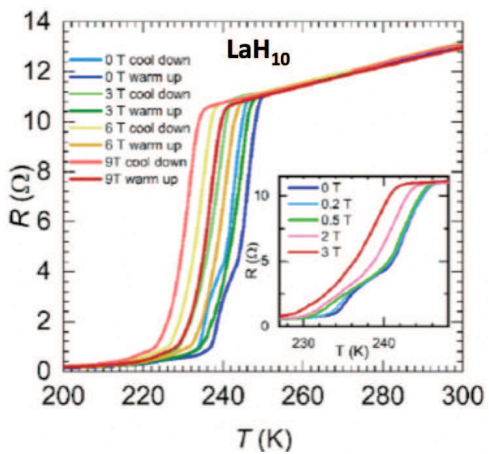}} 
 \caption {Resistive transition for $LaH_{10}$ in a magnetic field ranging from 0 to 9T, from
 A. P. Drozdov, P. P. Kong, V. S. Minkov, S. P. Besedin, M. A. Kuzovnikov, S. Mozaffari, L. Balicas, F. F. Balakirev, D. E. Graf,
     V. B. Prakapenka, E. Greenberg, D. A. Knyazev, M. Tkacz and M. I. Eremets, 
     \href{https://www.nature.com/articles/s41586-019-1201-8}{Nature 569, 528-531 (2019)}, Ref. \cite{eremetslah}.}
 \label{figure9}
 \end{figure} 
 
         \begin{figure} [t]
\resizebox{6.5cm}{!}{\includegraphics[width=6cm]{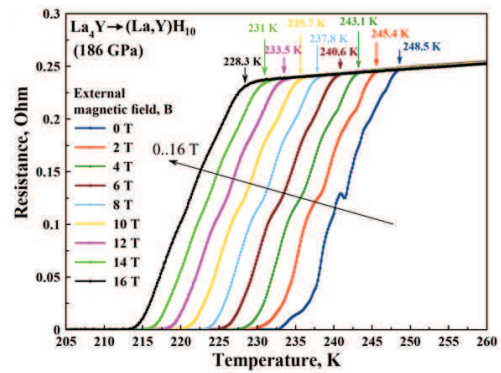}} 
 \caption {Resistive transition for $(La,Y)H_{10}$ in a magnetic field ranging from 0 to 16T, from
 Dmitrii V. Semenok, Ivan A. Troyan, Alexander G. Kvashnin, Anna G. Ivanova, Michael Hanfland, Andrey V. Sadakov, Oleg A. Sobolevskiy, Kirill S. Pervakov, Alexander G. Gavriliuk, Igor S. Lyubutin, Konstantin V. Glazyrin, Nico Giordano, Denis N. Karimov, Alexander B. Vasiliev, Ryosuke Akashi, Vladimir M. Pudalov, Artem R. Oganov,  \href{https://arxiv.org/abs/2012.04787}{arXiv:2012.04787 (2020)}, Ref. \cite{layh10}.}
 \label{figure10}
 \end{figure}

 In hydride superconductors, the resistive transition in a field looks qualitatively different to what it looks like in  standard superconductors,
 as shown in Figs. 6-10.
 
 The most glaring difference is seen for the case of the `room temperature superconductor' CSH \cite{roomt}, Fig. 6.
 The transition is already unusually sharp in the absence of applied magnetic field: $T_c$ drops to zero over a 
 range of temperature not larger than $\Delta T=0.5K$, hence a fractional broadening 
 $\Delta T/T=0.0018$. Such sharp transitions are not seen in standard superconductors except for exceptionally pure type I superconductors.
 In all the examples shown in Figs. 2-5, the width of the transition at $H=0$ is at least $\Delta T/T>0.02$, i.e. an order of magnitude
 larger. In addition, the transition in CSH  remains equally sharp upon application of a magnetic field as high as $9T$. Given the estimated critical
 field $H_{c2}=62T$ for this case, $H=9T$ is $15\%$ of $H_{c2}$.  
 
 For the other examples of hydride superconductors shown (Figs. 7, 8, 9, 10) the transition in the absence of a field is broader, more in line with
 standard superconductors, but the behavior under application of magnetic field is equally anomalous. Either the
 transition width doesn't change with field, or it even narrows, as seen in Fig. 7 for $YH_9$.

         \begin{figure} [t]
 \resizebox{7.5cm}{!}{\includegraphics[width=6cm]{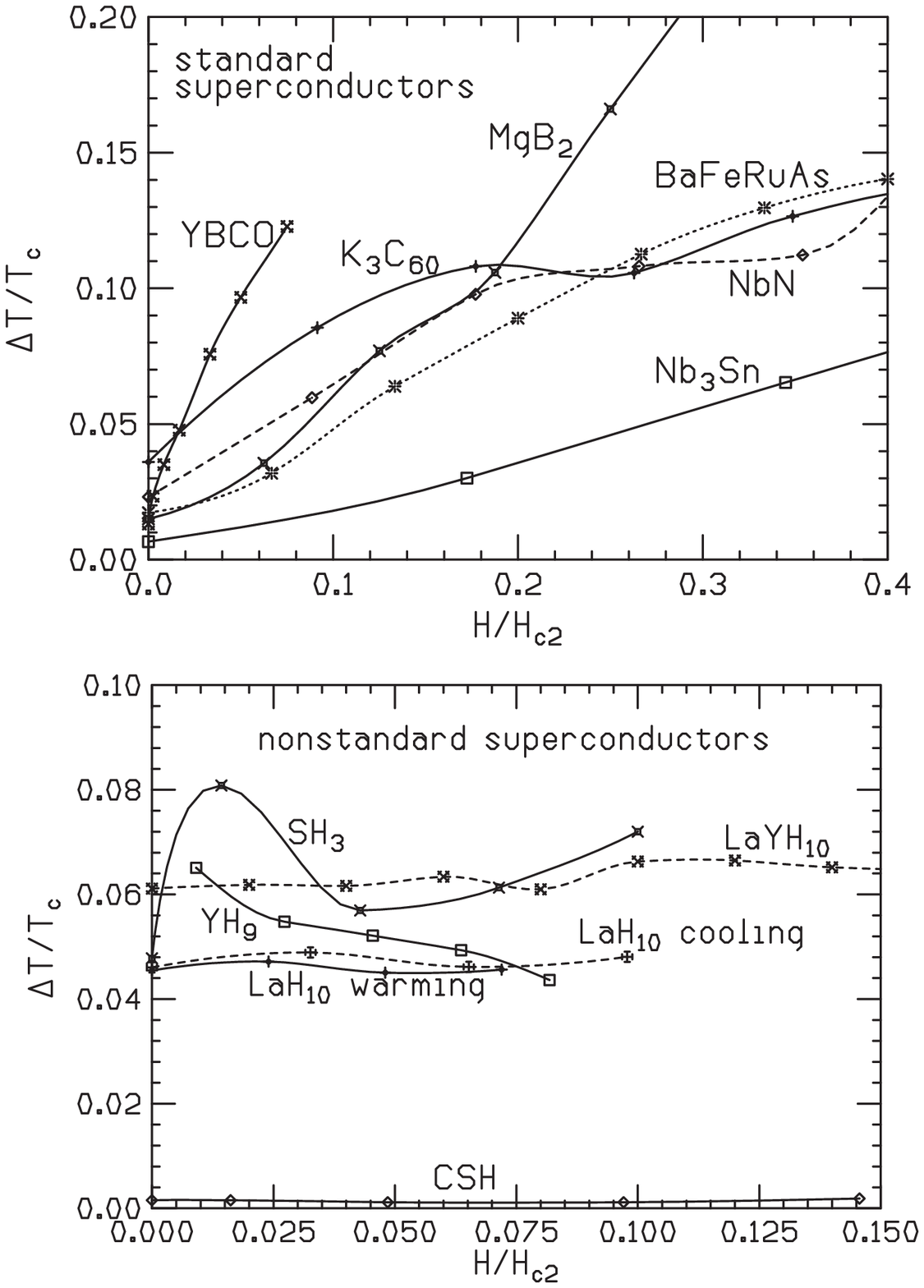}} 
 \caption { Upper panel: YBCO single crystal, $H_{c2}=120$~T, $T_c^{max}=91K$ \cite{iye}; 
 $MgB_2$ polycrystalline, $H_{c2}=16T$, $T_c^{max}=40K$ \cite{canfield}; 
  $K_3C_{60}$ single crystal, $H_{c2}=17.5T$, $T_c^{max}=20K$ \cite{k3c60}; 
   $NbN$ thin film epitaxial, $H_{c2}=11.3T$, $T_c^{max}=17K$ \cite{nbn}; 
    $Nb_3Sn$ polycrystalline,  $H_{c2}=29$ T, $T_c^{max}=18K$ \cite{nb3sn}; 
       $BaFe_{2-x}Ru_xAs_2$ $x=0.71$,  single crystal, $H_{c2}=30T$, $T_c^{max}=20K$ \cite{bafe}.
       Lower panel: 
        $SH_3$, $P=155 GPa$,  $H_{c2}=70T$, $T_c^{max}=180K$ \cite{eremetssh}; 
         $YH_9$, $P=185 GPa$,  $H_{c2}=110T$, $T_c^{max}=218K$ \cite{yttrium2}; 
         $LaH_{10}$ cooling, $P=150 GPa$,  $H_{c2}=92T$, $T_c^{max}=245K$ \cite{eremetslah}; 
         $LaH_{10}$ warming, $P=150 GPa$,  $H_{c2}=125T$, $T_c^{max}=249K$ \cite{eremetslah}; 
         $CSH$, $P=267 GPa$,  $H_{c2}=61.8T$, $T_c^{max}=288K$ \cite{roomt},
          $(La,Y)H_{10}$, $P=186 GPa$,  $H_{c2}=100T$, $T_c^{max}=248.5K$ \cite{layh10}.}
 \label{figure11}
 \end{figure} 
 
  In Fig.~\ref{figure11} we contrast the qualitatively different behavior of standard and nonstandard superconductors, showing
  several representative examples.
 The horizontal axis gives $H$, the applied field, divided by the reported upper critical field for each case. 
 The vertical axis gives the ratio of the broadening of the transition $\Delta T$ divided by the temperature $T_c$ where
 the resistance starts to drop for each applied magnetic field.
 
 For all standard superconductors shown in Fig.~\ref{figure11} top panel, the transition broadens under application of a magnetic field,
 as expected from the theory discussed in the next section. We have looked at many more examples of standard
 superconductors and the qualitative behavior is always the same. Note that the upward slope is larger for the 
 higher $T_c$ cases, $YBCO$ (unconventional) and $MgB_2$ (conventional), as expected from the  standard theory discussed in the
 next section. This  leads to the expectation that the slope
 should be even larger for the  high $T_c$ hydrides under pressure. Nothing of the sort is seen in the lower panel of
 Fig.~\ref{figure11}.
 
 Instead, we see in the lower panel of Fig.~\ref{figure11} that there is no upward trend in the broadening of the transition
 with increased magnetic field. Quite the contrary, the broadening either stays the same as for $H=0$ or even decreases
 with applied field, as the data for $YH_9$ show. Note also that the broadening for CSH, both without magnetic field
 and with an applied magnetic field of up to $15\%$ of the upper critical field, is two orders of magnitude smaller
 than any of the broadening seen in the top panel for standard superconductors.
 The data in Fig.~\ref{figure11} clearly show that high temperature hydride superconductors are qualitatively different from standard
 superconductors as far as the behavior of the resistance versus temperature in an applied magnetic field is concerned.
 This cannot be accounted for within the standard theory of superconductivity, as discussed in the next section.
 
  \section{resistive broadening under an applied magnetic field}
  
  As mentioned above, when a type II superconductor is cooled in the presence of an applied
  magnetic field $H$, it undergoes a second order phase transition to a mixed state at temperature $T_c(H)$ that satisfies
  $H_{c2}(T_c(H))=H$. Immediately below this temperature, the normal vortex cores in this mixed phase occupy approximately
  half of the volume of the material according to Eq. (4b). When current flows, vortices are expected to  flow driven by the Lorentz force 
  which is equivalent to the flow of normal current, giving rise to a finite voltage drop across the electrodes and Joule heat dissipation.
  Hence there is non-zero resistance, called flux-flow resistance.
  As the temperature is lowered further for fixed magnetic field, the density of vortices stays constant but their core size shrinks, the condensation energy
  increases and a larger 
  fraction of the material enters the superconducting state. At some lower temperature it is observed that the resistance drops to zero or to
  immeasurably small values, which is attributed to the vortices becoming pinned at impurities or imperfections in the sample, preventing their motion.
  The strength of pinning potential needed to pin the vortices also depends on the interaction between vortices:
  if the interaction is sufficiently large that they become rigidly connected to each other forming a `vortex glass' \cite{fisher}, it becomes easier to pin
  the entire vortex assembly than if the vortices are pinned independently.
  
  Why should we expect the broadening of the transition to be larger for higher $T_c$ materials, and to increase with  increased magnetic field? 
  We can understand it  as follows. As discussed by Anderson \cite{anderson} and Anderson and Kim \cite{andersonkim},
  let us call $U_0$ the characteristic energy of pinning. Thermally activated depinning will be proportional to 
  $e^{-U_0/(k_BT)}$, so for given pinning energy it will be exponentially larger at higher temperature. For a given vortex to remain stationary,
  the pinning threshold energy is given approximately by \cite{anderson}
  \beq
  U_0 \sim \frac{H_c}{8\pi} \xi^3,
  \eeq
  with $H_c$ the thermodynamic critical field. 
  This indicates that the required pinning energy becomes smaller for shorter coherence length; this is the case for more strongly type II superconductors. 
  
  In addition, Yeshurun and Malozemoff argued \cite{yeshurun88} that when the flux lattice spacing  (distance between nearest neighbor vortices) $a_0\sim (\phi_0/H)^{1/2}$,
  with $H$ the applied magnetic field,  becomes significantly smaller than the
  London penetration depth, a cross-over to collective pinning should occur where 
  \beq
  U_0\sim \frac{H_c^2}{8\pi}a_0^2\xi .
  \label{U0_YM}
  \eeq
  Within GL theory Eqs.~(\ref{hc_gl}) and (\ref{xi_gl}) give $H_c\sim H_{c0}(1-t)$, $\xi \sim \xi_0(1-t)^{-1/2}$, with $t=T/T_c$ the reduced temperature, and hence Eq.~(\ref{U0_YM}) yields
  \beq
  U_0=C \frac{H_{c0}^2}{8\pi}\phi_0 \xi_0 \frac{(1-t)^{3/2}}{H}
   \label{U0_YM_2}
  \eeq
  where C is a numerical factor. The pinning energy decreases with increased magnetic field $H$, which gives larger broadening for larger $H$.
  
 This realization provided some theoretical support for the work of  M\"uller {\it et al.}  \cite{muller87}, who recognized almost  immediately upon discovering cuprate superconductivity
 that an ``irreversibility line'' exists in the $H-T$ plane. This line delineates the reversible region from the
region where magnetization shows hysteresis. They also found that the irreversibility line had a temperature dependence,
$H_{\rm irr} \propto (T_c - T)^{3/2}$. Tinkham \cite{tinkham88} 
then made use of this temperature dependence attributed to the
activation energy for flux motion \cite{yeshurun88} 
to understand the resistance in this regime, following the Ambegaokar-Halperin \cite{ambegaokar69}
theory of the resistance due to thermally activated phase slippage. From Eq.~(\ref{U0_YM_2}), Tinkham argued that the flux pinning barrier height $U_0$ normalized to
$k_BT$ would be given by an expression, 
\begin{equation}
\gamma_0 \equiv {U_0 \over k_BT} = C^\prime {J_{c0}(0) \over T_c} {(1-t)^{3/2} \over H},
\label{gamma0}
\end{equation}
where $C^\prime$ is a constant, and $J_{c0}(0)$ is the critical current density for $T=0$ and $H=0$. Insertion into the Ambegaokar-Halperin
expression \cite{ambegaokar69} results in the Tinkham formula
\begin{equation}
{R \over R_n} = {1 \over I_0^2[A (1-t)^{3/2}/(2H/H_{c2})]},
\label{tinkham}
\end{equation}
where $A$ is a dimensionless constant, $R_n$ is the normal state resistance and $I_0(x)$ is the modified Bessel function of the first kind.
As Tinkham illustrated \cite{tinkham88} with the data of Ref.~[\onlinecite{iye}] (also shown in our Fig.~\ref{figure3}), 
this formula reproduces these curves in
considerable detail with one fitting parameter, $A=10$. As noted by Tinkham \cite{tinkham88}, his fitted value of this constant 
is the correct order of magnitude
expected for the YBCO, given the measured value of $J_{c0}(0)$ and the extrapolated value of $H_{c2}$. This is also true for the other materials
(see below) shown in the upper panel of Fig.~\ref{figure11}.
For our purposes, the important outcome is that for a type II superconductor, the width of
the transition, $\Delta T$, naturally increases with increasing applied field. For the cuprates, following \cite{tinkham88}, one finds
\begin{equation}
\Delta T \propto H^{2/3}.
\label{tinkham_scale}
\end{equation}

\begin{figure} [t]
 \resizebox{8.5cm}{!}{\includegraphics[width=6cm]{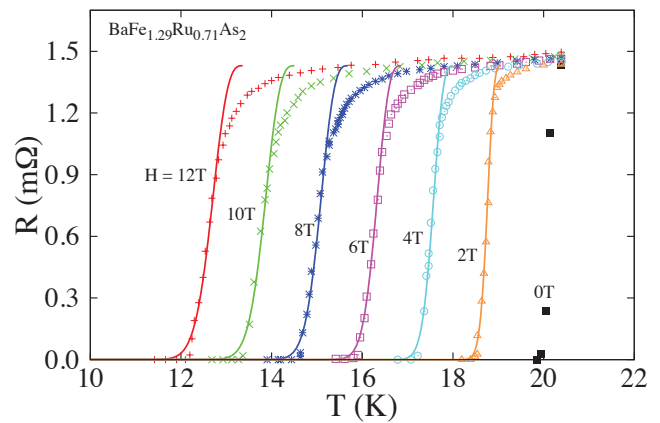}} 
 \caption{ Resistance vs temperature for BaFe$_{1.29}$Ru$_{0.71}$As$_2$ for a variety of applied magnetic fields, as indicated. Experimental
 data, represented by the points for the various fields, is taken from Ref.~\cite{bafe}. Our theoretical fits are given by curves. We normalized the
 fits to the normal state resistance at $T_c = 20.3$ K, and used the single fitting parameter, $A = 83.3$, as discussed in the text. We used
 $T_c(H) = 19.05, 18.0, 16.85, 15.68, 14.5 13.36, 12.0$ K for $H = 2, 4, 6, 8, 10, 12$ T, respectively. These form a linear dependence of $H_{c2}(T)$
 and extrapolate to $H_{c2}(0) = 25$ T (WHH theory \cite{whh}) or $35$ T (two-fluid model).}
 \label{figure12}
 \end{figure}

The broadening that occurs in the Iye data \cite{iye} is visually obvious. We showed other systems that clearly exhibit similar broadening.
Even where it is not so obvious, for example in the pnictides \cite{bafe} (see their Fig.~1a), the Tinkham formula
works very well in reproducing the broadening trend illustrated in Fig.~\ref{figure12}. 
We have generalized this formula slightly by using $t \equiv T/T_c(H)$.
An $H$-dependent $T_c$ was not required in the case of YBCO. The result is shown in Fig.~\ref{figure12}, 
again with a single fitting parameter, $A=83.3$ in
this case. We actually used an optimal fit for $T_c(H)$ in obtaining the curves shown, but when plotting the resulting $H_{c2}$ vs. $T_c$, we obtain
a straight line near $T_{c0} \equiv T_c(H=0)$, as expected in the two-fluid model (or any other more sophisticated model). We could have started
with this curve, dictated by experiment, so, for this reason we do not regard these as fitting parameters. In fact we consider this self-consistency
a further check of the theory.
 Note that a significant amount of broadening occurs, and the Tinkham formula captures the majority of this broadening. As is clear from
 Fig.~\ref{figure12} there is an additional ``rounding'' of the resistivity that the Tinkham formula does not capture, but nonetheless increases
 with increasing applied magnetic field. We do not pursue the origin of this additional rounding here.

 The previous expression assumes that the intrinsic
 width, with no applied magnetic field, $H_0$, is zero, but in general this is not the case for a variety of possible reasons
 such as disorder in the sample.
 To take into account this intrinsic width we substitute in Tinkham's formula Eq. (\ref{tinkham}) $1-t=\Delta T/T_c$ by 
 $\Delta T/T_c-\Delta T_0/T_c$, where $\Delta T_0$ is the intrinsic broadening for $H=0$.
 The behavior seen in all the examples shown in Fig.~\ref{figure11} upper panel can then be fitted semi-quantitatively
 with the parameter
 $A$ in Eq. (26) ranging from  10 for  $YBCO$  to 257 for $Nb_3Sn$
 (A=42.4, 45.1, 38.9 for $MgB_2$, $NbN$, $K_3C_{60}$ respectively). Instead, the behavior shown in the lower panel of Fig.~\ref{figure11} 
 for CSH cannot be fitted to the Tinkham formula Eq. (26) unless 
 $A > 100,000$. Given the values of $T_c$ and $H_{c2}$ for CSH, then, following Tinkham's analysis for YBCO \cite{tinkham88}, this implies
 a zero temperature critical current density in excess of $10^{11}$ A/cm$^2$ ! The occurrence of such a large critical current density should be easily
 tested by experiment.
 For $YH_9$ the fitting fails completely since the width of the transition decreases instead
 of increasing with magnetic field.
 We conclude that if hydrides under pressure are truly superconductors,   they do not
 obey the same physical principles that all other type II superconductors, whether conventional or unconventional, obey. 
 
 \section{magnetic measurements}

           \begin{figure*} [t]
 \resizebox{16.5cm}{!}{\includegraphics[width=18cm]{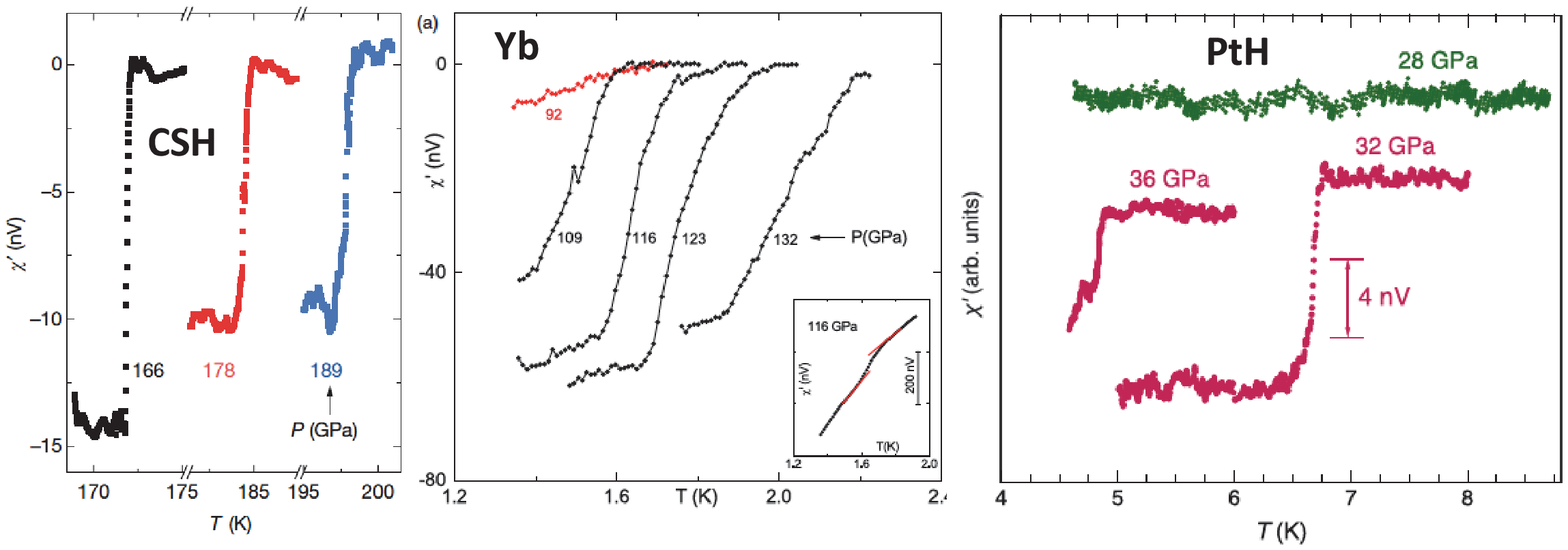}} 
 \caption { Ac magnetic susceptibility data for materials under pressure. Left panel: CSH, from
 Elliot Snider, Nathan Dasenbrock-Gammon, Raymond McBride, Mathew Debessai, Hiranya Vindana, 
     Kevin Vencatasamy, Keith V. Lawler, Ashkan Salamat and Ranga P. Dias,
     \href{https://www.nature.com/articles/s41586-020-2801-z}{Nature 586, 373 (2020)}, Ref. \cite{roomt}; middle panel: Yb, from 
    J. Song, G. Fabbris, W. Bi, D. Haskel, and J.S. Schilling, 
    \href{https://journals.aps.org/prl/abstract/10.1103/PhysRevLett.121.037004}{Phys. Rev. Lett. 121, 037004 (2018)}, Ref. \cite{yb};
    right panel: PtH, from
    Takahiro Matsuoka, Masahiro Hishida, Keiji Kuno, Naohisa Hirao, Yasuo Ohishi, Shigeo Sasaki, Kazushi Takahama,
and Katsuya Shimizu, \href{https://journals.aps.org/prb/abstract/10.1103/PhysRevB.99.144511}{Phys. Rev. B 99, 144511 (2019)}, Ref. \cite{pth}. 
 The width of the transition is $\Delta T/T_c\sim 0.03$ for $PtH$ ($P=32 G Pa$), $\Delta T/T_c\sim 0.11$ for $Yb$ ($P=116 G Pa$) and $\Delta T/T_c\sim 0.001$ for $CSH$ ($P=166 G Pa$)}
 \label{figure13}
 \end{figure*}

 In addition to transport measurements, some of the experimental papers on high temperature hydride superconductors also report results of magnetic measurements that appear to support the claim of superconductivity. Let us consider those measurements.
 Reference \cite{roomt} shows results for  ac magnetic susceptibility measurements in CSH  that indicate a sharp drop at certain temperatures,
 that the authors claim is consistent with onset of superconductivity. First, it should be pointed out that the magnetic measurements are not done
 during the same run where the transport data are collected due to experimental constraints, and as a consequence it is not possible to verify
 whether the signals indicating superconductivity show up at the same temperatures in transport and magnetic data.
 In fact, Ref. \cite{roomt} comments that the susceptibility signal occurs at somewhat higher temperature than the transport signal,
 which they attribute to uncertainties in the estimated pressures due to pressure gradients.
 Furthermore, the transitions shown in the susceptibility data are also anomalously sharp. 
 For example, the width of the transition for 166 GPa is merely $\Delta T=0.2K$, hence $\Delta T/T_c=0.001$.
 Typical  ac susceptibility measurements for materials under pressure in diamond anvil cells  show much broader transitions \cite{back,tim,hamlinthesis,md,yb,pth}.
 We show some examples in Fig.~\ref{figure13}. Given that the magnetic penetration depth is much larger than the 
 coherence length in CSH and other hydrides,
 and that half of the sample remains in the normal state at the onset of superconductivity, one would expect a much broader transition in the magnetic susceptibility than seen in the left panel of Fig.~\ref{figure13}. The same is true for ac magnetic susceptibility data for other hydrides \cite{chish3}.
Other reasons for why the ac magnetic susceptibility measurements on CSH may not indicate superconductivity were
 discussed in Ref. \cite{magnetic}.


 In Ref. \cite{nrs}, it has been claimed that the Meissner effect was detected in 
sulfur hydride  through a novel nuclear resonant scattering experiment, 
and hence that the existence of superconductivity has been unequivocally confirmed \cite{nrs,nrs2}.
However we have challenged that claim  in Ref. \cite{meissnerours}. We showed that a standard superconductor, whether conventional or
unconventional, would not show the behavior reported in Ref. \cite{nrs}. 
More specifically, that in order to show that behavior the critical current would have to be enormous, qualitatively larger than for standard superconductors,
similarly to what we found in this paper fron consideration of the width of the resistive transition. 
We concluded in \cite{meissnerours} that   either the observations reported in Ref. \cite{nrs} were experimental artifacts and hence provided no evidence for superconductivity, or
alternatively that they provided further evidence that 
hydride superconductors are  nonstandard superconductors, fundamentally different from standard superconductors.

\section{Discussion}
  
In this paper we have considered the question of whether superconductivity in hydrides under high pressure follows the expected behavior of other superconductors.
We have found that the answer is negative.

Our finding is remarkable because hydride superconductors are considered to be textbook examples of 
conventional BCS-Eliashberg superconductors driven by the electron-phonon interaction, which unlike unconventional superconductivity has been thought to be
well understood for the last 60 years. All the theoretical analysis of these materials
conclude that their $T_c$ is perfectly well explained by standard superconductivity theory
 \cite{th0,th1,th2,th3,th4,th5,th6,th7,th8,th9,th10,th11,th12}. How is it possible then that 
the behavior of resistivity in the presence of a magnetic field does not follow the standard behavior seen in type II superconductors \cite{tinkham}?
How is it possible that magnetic measurements also show anomalous behavior incompatible with standard superconductivity?

The only conceivable explanation within the standard theory for the absence of broadening of the resistive transitions in a magnetic field  would be that for some reason there is an enormous pinning potential that confines the vortices 
immediately as the system enters the superconducting state, much larger than in standard superconductors. If that was the case,
it would also be true that the critical current should be enormous in these superconductors, opening up the way for important technological applications. Unfortunately we know of no physical reason why 
these materials should have such anomalously large pinning potentials.  Furthermore, even if that was the case  it would not explain  the extreme sharpness of the transition in magnetic susceptibility measurements, which is not affected by the strength of the pinning potential.

We conclude then that there are two possibilities:

(1) These materials constitute a new category, `nonstandard superconductors', exhibiting physics that is qualitatively different from that of  standard conventional and unconventional 
superconductors. 
Perhaps these materials are simultaneously type I and strongly type II superconductors, exhibiting a new kind of `complementarity'.
Recall that for standard  type I superconductors the transition to superconductivity in a field is first order, while for type II it is second order.
Perhaps even the classification of type I versus type II does not apply to them. 
Perhaps the coherence length and the London penetration
depth are not the relevant lengths. Their electrodynamic behavior needs to be understood in a new theoretical framework, unlike the one formulated for
standard superconductors. If this is the case it is very remarkable that the conventional framework of BCS-Eliashberg theory would still apply to them
with no modification, as found in the theoretical analyses \cite{th0,th1,th2,th3,th4,th5,th6,th7,th8,th9,th10,th11,th12}.

 (2) These materials are not superconductors. Their measured properties that have been interpreted as reflecting superconductivity are instead reflecting
either (a) experimental artifacts or  (b) different physics, or (c) a combination of both. Let us consider these possibilities.

The possibility that experimental artifacts may be responsible for all or some of the signals mistakenly interpreted as superconductivity cannot be ruled out a priori
in any experiment. Examples in the scientific literature where claims of superconductivity were subsequently withdrawn abound.
The strong desire to find superconductivity in a compound may cause careful researchers to misinterpret the origin of observations.
We feel that research in high pressure hydride superconductivity may have reached the stage where theoretical bias is unduly guiding the interpretation of
experimental findings.
\cite{ashcroft,ashcroft2,review1,review2,review3}

With respect to different physics, quite generally the sharpness of the signals observed suggests that if they indicate phase transitions
they are first rather than second order transitions. We suggested in \cite{hm} that the abrupt increases in resistance found upon heating may result from 
metallic paths being destroyed as the temperature is increased. This could result from  atomic rearrangements due to a local first order phase transition
between phases that are close in free energy but have very different electrical conductivities.  This was also recently suggested by Dogan and Cohen \cite{cohen}. The fact that the transitions shift  to lower temperature upon application
of a magnetic field could result from the fact that as the magnetic field is applied, Joule heating of the sample produced by
eddy currents may randomly rearrange atomic positions. Cases where such processes may result in increase rather than 
decrease of `critical temperature' may be disregarded as spurious because they don't conform to the expected behavior
predicted by theory.

Alternatively, we have suggested in Ref. \cite{magnetic} that other different physics could be
the onset of  weak ferromagnetic order. This is supported by the raw data in susceptibility measurements reported in Ref. \cite{roomt},
and would be in agreement with alternative theoretical expectations \cite{fm}. Refs. \cite{mazov}  and  \cite{cohen} also suggested that magnetic effects may
account for some of the observations, and they also suggested alternative possibilities not discussed here.

In conclusion, more experimental and theoretical work is needed to establish what is going on in these materials.

\begin{acknowledgments}
FM 
was supported in part by the Natural Sciences and Engineering
Research Council of Canada (NSERC) and by an MIF from the Province of Alberta.

\end{acknowledgments}

 \end{document}